\documentclass[fleqn,10pt]{wlscirep}
\usepackage[utf8]{inputenc}
\usepackage[T1]{fontenc}

\usepackage{bm}
\usepackage{xcolor}

\title{Deforestation and world  population sustainability: a quantitative analysis }

\author[1]{Mauro Bologna}
\author[2,3,4*]{Gerardo Aquino}
\affil[1]{Departamento de Ingenier\'ia El\'ectrica-Electr\'onica, Universidad de Tarapac\'a}
\affil[2]{The Alan Turing Institute, London, UK}
\affil[3]{University of Surrey, Guildford, UK.}
\affil[4]{Goldsmiths, University of London, UK}
\affil[*]{gaquino@turing.ac.uk}

\affil[+]{these authors contributed equally to this work}

\begin{abstract}
In this paper we afford a quantitative analysis of the sustainability of current world population growth in relation to the parallel deforestation process adopting a statistical point of view. We consider a simplified model based on a stochastic growth process driven by a continuous time random walk, which depicts  the technological evolution of human kind,  in conjunction with  a deterministic  generalised logistic model for humans-forest interaction and  we evaluate the probability of avoiding the self-destruction of our civilisation. {\color{black}Based on the current resource consumption rates and best estimate of technological rate growth}  our study shows that we have very low probability, less than $10\% $ in most optimistic estimate, to survive without facing a catastrophic collapse. 

\end{abstract}
\begin{document}

\flushbottom
\maketitle
\thispagestyle{empty}

\noindent

\section*{Introduction}

In the last few decades, the debate on climate change has assumed global importance with consequences on national and global policies. Many factors  due to human activity are considered  as possible responsible of the observed changes:  among these water and air contamination (mostly greenhouse effect) and deforestation are the mostly cited. While the extent of human contribution to the greenhouse effect  and temperature changes is still  a matter of discussion, the deforestation is an undeniable fact. Indeed before the development of human civilisations,  our planet was covered by 60 million square kilometres of forest~\cite{rich}.  As a result of deforestation, less than 40 million square kilometres  currently remain~\cite{fao}. In this paper, we focus on the consequence of indiscriminate deforestation.

Trees' services to our planet range from carbon storage, oxygen production to soil conservation and water cycle regulation. They support natural and human food systems and provide homes for countless species, including us, through building materials. Trees and forests are  our best atmosphere cleaners and, due to the key role they play in the terrestrial ecosystem, it is highly unlikely to imagine the survival of many species, including ours,  on Earth without them. In this sense, the debate on climate change will be almost obsolete in case of a global deforestation of the planet. 

Starting from this almost obvious observation, we investigate the problem of the survival of humanity from a statistical point of view. We   model  the interaction between forests and humans based on a deterministic logistic-like dynamics, while we assume a stochastic model for the technological development of the human civilisation. The former model has already been applied in similar contexts\cite{epl, mauro2} while the latter is based on data and model of  global energy consumption \cite{elov,emeritus} used as a proxy for the technological development of a society. This gives solidity to our discussion and we show that, keeping the current rate of deforestation, statistically the probability to survive without facing a catastrophic collapse, is very low. We connect such probability to survive to the capability of humankind  to spread and exploit the resources of the full solar system. According to Kardashev scale~\cite{nic,nic2},  which measures a civilisation's level of technological advancement based on the amount of energy they are able to use, in order to spread  through the solar system we need to be able to harness the energy radiated by the Sun  at a rate of ~$\approx 4\times10^{26}$ Watt. Our current energy consumption rate is estimated in~$\approx 10^{13}$ Watt~\cite{srw}. As showed in the subsections  "Statistical Model of technological development" and "Numerical results" of the following section,  a successful outcome has a well defined threshold and we conclude that the probability of avoiding  a catastrophic collapse is very low, less than $10 \%$ in the most optimistic estimate.

\section*{Model and Results}
\subsection*{Deforestation}
The deforestation of the planet is a fact~\cite{fao}. Between 2000 and 2012, 2.3 million Km$^2$ of forests around the world were cut down~\cite{wiki} which amounts to~$2\times 10^5$ Km$^2$  per year.  At this rate all the forests would disappear approximatively in $100-200$ years. Clearly it is unrealistic to imagine that the human society would start to be affected by the deforestation only when the last tree would be cut down. The progressive degradation of the environment due to deforestation would heavily affect human society and consequently the human collapse would start much earlier.

Curiously enough, the current situation of our planet  has a lot in common with the deforestation of Easter Island as described in ~\cite{epl}.  We therefore use the model introduced in that reference to roughly describe the  humans-forest interaction. Admittedly, we are not aiming here for an exact exhaustive model. It is probably impossible to build such a model. What we propose  and illustrate in the following sections, is a simplified model which nonetheless allows us to extrapolate the time scales of  the processes involved: i.e. the deterministic  process describing human population and resource (forest) consumption  and the stochastic process defining the economic and technological growth of societies. Adopting the model in \cite{epl} (see also \cite{frank}) we have for the humans-forest dynamics

\begin{eqnarray}\label{res-inter2}
\frac{d}{dt}N(t)&=&rN(t)\left[ 1-\frac{N(t)}{\beta R(t)}\right] ,
\\\label{log-inter2}
\frac{d}{dt}R(t)&=&r^{\prime }R(t)\left[ 1-\frac{R(t)}{R_{c}}
\right] -a_0 N(t)R(t) .
\end{eqnarray}
where~$N$ represent the world population and~$R$ the Earth surface covered by forest.  $\beta$ is a positive constant related to the  carrying capacity of the planet  for human population, $r$ is the growth rate for humans (estimated as $r\sim 0.01$years$^{-1}$)~\cite{fort}, $a_0$ may be identified as the technological parameter measuring  the rate at  which humans can  extract the resources from the environment,  as a consequence of their reached technological level.  $r'$ is the renewability parameter representing the capability of the resources to regenerate, (estimated as $r' \sim 0.001$ years$^{-1}$)\cite{porcomauro} , $R_c$ the resources carrying capacity that in our case may be identified with the initial 60 million square kilometres of forest. 

A closer look at this simplified  model and  at  the analogy with Easter Island on which is based,  shows  nonetheless,   strong similarities with our current situation.   Like the old inhabitants of Easter Island  we too,  at least for few more decades, cannot leave the planet. The consumption of the natural resources, in particular the forests, is in competition with our technological level.  Higher technological level leads to growing population and higher forest consumption {\color{black}(larger $a_0$)} but also to a more effective use of resources. With higher  technological level we can in principle develop technical solutions to avoid/prevent the ecological collapse of our planet or, as last chance,  to rebuild a civilisation in the extraterrestrial space {\color{black}( see  section on the Fermi paradox)}. The  dynamics  of our model for  humans-forest interaction in Eqs. (\ref{res-inter2} , \ref{log-inter2}), is typically characterised by a growing human population until a maximum is reached after which a rapid disastrous collapse in population occurs before eventually reaching a low population  steady state or  total extinction. We will use this maximum as a reference for reaching a disastrous condition. We call this point in time the "no-return point" because if the deforestation rate is not changed before this time the human population will not be able to sustain itself and a disastrous collapse or even extinction  will occur. 
As a  first approximation~\cite{epl}, since the capability of the resources to regenerate, $r'$ , {\color{black} is an order of magnitude }smaller than the growing rate for humans, $r$, we may neglect the first term in the right hand-side of Eq.~(\ref{log-inter2}).  Therefore, working in a regime of the exploitation of the resources governed essentially by the deforestation, from Eq.~(\ref{log-inter2}) we  can derive  the rate of tree extinction as

\begin{equation}
\frac{1}{R}\frac{dR}{dt}\approx -a_0 N.
\label{ext}
\end{equation}
The actual population of the Earth is $N\sim 7.5 \times 10^{9}$ inhabitants with a maximum carrying capacity estimated~\cite{wil} of $N_c\sim  10^{10}$ inhabitants. The forest carrying capacity may be taken as~\cite{rich} $R_c\sim 6\times 10^7$ Km$^2$ while the actual surface of forest is $R\lesssim4\times 10^7$ Km$^2$. Assuming that $\beta$ is constant, we may estimate this parameter evaluating the equality $N_c(t)=\beta R(t)$ at the time when the forests were intact. Here $N_c(t)$ is the instantaneous human carrying capacity given by Eq.~(\ref{res-inter2}). We obtain $\beta \sim N_c/R_c\sim 170$. 

\begin{figure}[h!]
\centering
\includegraphics[width=.45\textwidth]{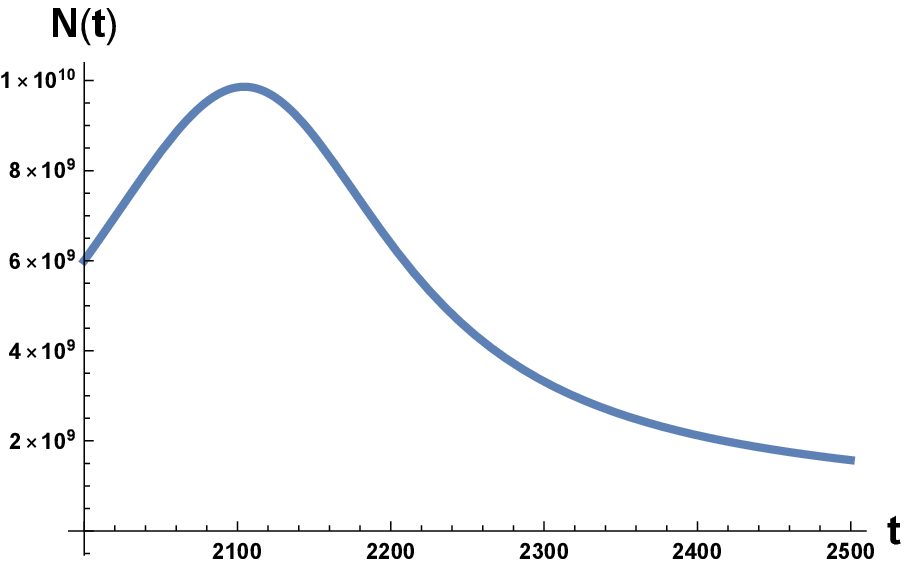}
\includegraphics[width=.45\textwidth]{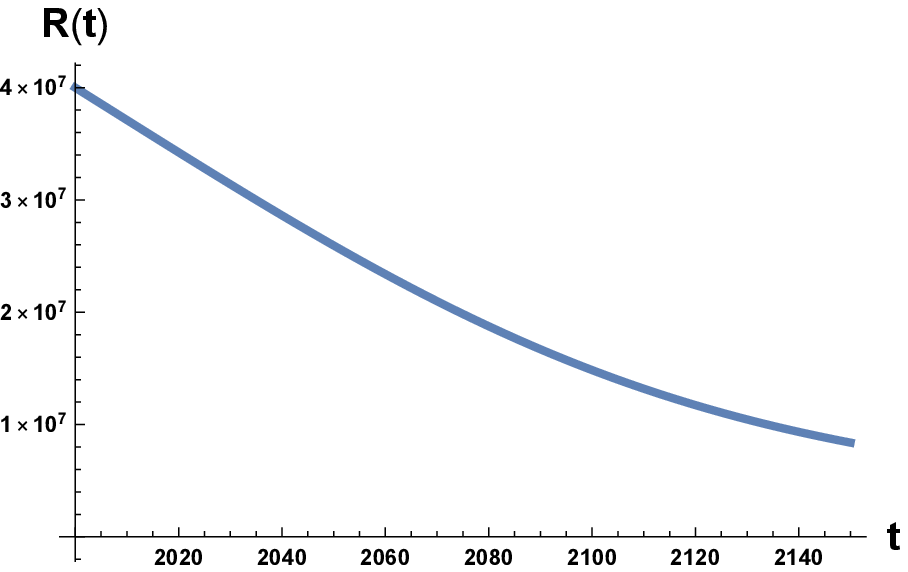}
\caption{On the left: plot of the solution of Eq. (\ref{res-inter2}) with the initial condition $N_0=6\times 10^9$ at initial time $t=2000$ A.C.
On the right: plot of the solution of Eq.~(\ref{log-inter2}) with the initial condition $R_0=4\times 10^7$. Here $\beta=700$ and $a_0 =10^{-12}$
\label{fig1}}
\end{figure}

\begin{figure}[h!]
\centering
\includegraphics[width=.45\textwidth]{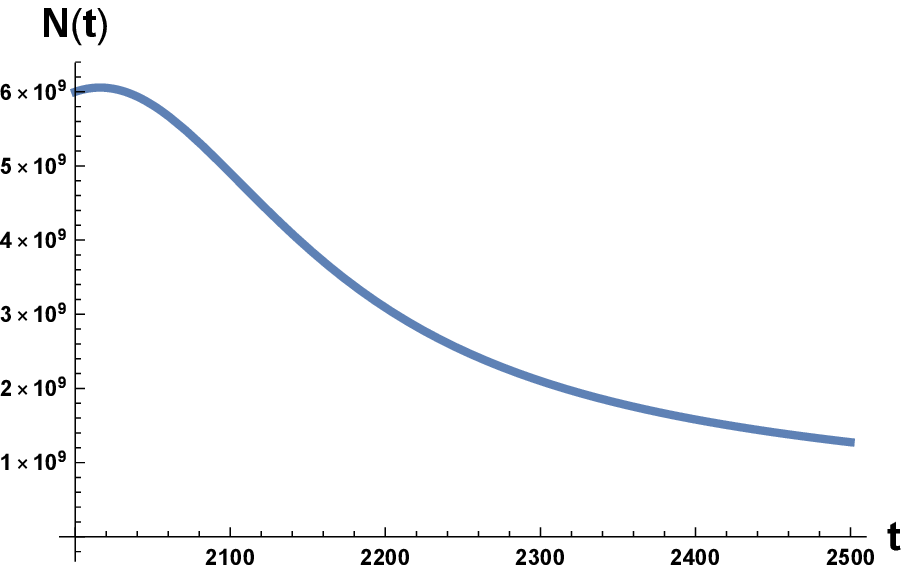}
\includegraphics[width=.45\textwidth]{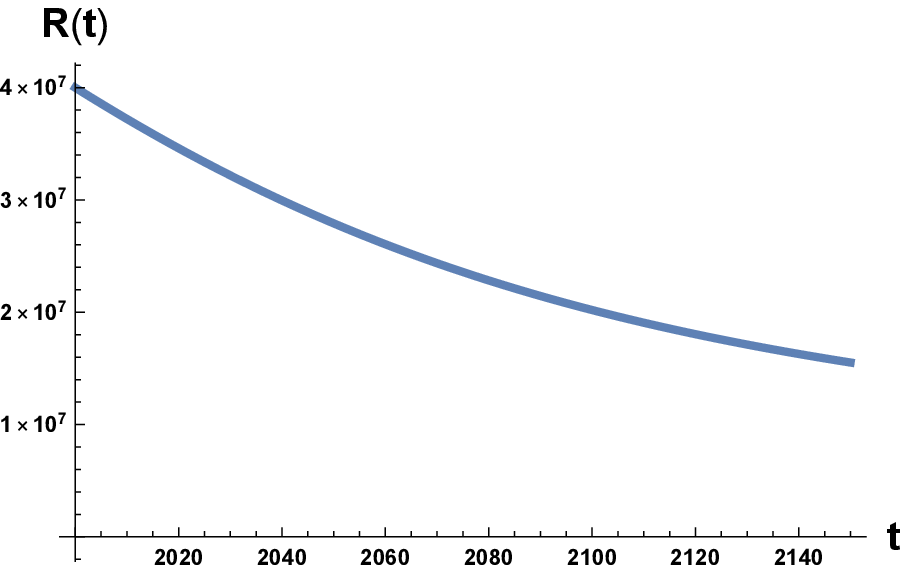}
\caption{On the left: plot of the solution of Eq.~(\ref{res-inter2}) with the initial condition $N_0=6\times 10^9$ at initial time $t=2000$ A.C.
On the right: plot of the solution of Eq.~(\ref{log-inter2}) with the initial condition $R_0=4\times 10^7$. Here $\beta=170$ and $a_0 =10^{-12}$
\label{fig2}}
\end{figure}

In alternative we may evaluate $\beta$ using actual data of the population growth~\cite{john} and inserting it in Eq.~(\ref{res-inter2}). In this case we obtain a range $700\lesssim\beta\lesssim 900$ that gives a slightly favourable scenario for the human kind (see below and Fig. \ref{fig4}). We stress anyway that this second scenario  depends on many factors not least the fact that the period examined in ~\cite{john} is relatively short. On the contrary $\beta\sim 170$ is based on the accepted value for the maximum human carrying capacity. With respect to the value of parameter $a_0$, adopting the data relative to years 2000-2012 of Ref. \cite{wiki}, we have

\begin{equation}
 \frac{1}{R}\frac{\Delta R}{\Delta t} \approx  \frac{1}{3\times 10^{7}}\frac{2.3 \times 10^6}{12} \approx -a_0 N
  \Rightarrow a_0\sim 10^{-12} \,\mathrm{years} ^{-1}
\label{ext_b}
\end{equation}
The time evolution of system (\ref{res-inter2})-(\ref{log-inter2}) is plotted in Figs.~\ref{fig1} and~\ref{fig2}. We note that in Fig.~\ref{fig1} the numerical value of the maximum of the function $N(t)$ is $N_M\sim  10^{10}$  estimated as the carrying capacity for the Earth population \cite{wil}. Again we have to stress that it is unrealistic to think that the decline of the population in a situation of strong environmental degradation would be a non-chaotic and well-ordered decline, that is also way we take the maximum in population and the time at which occurs as the point of reference for the occurrence of an irreversible catastrophic collapse, namely a 'no-return' point.

\subsection*{Statistical Model of technological development}
According to Kardashev scale~\cite{nic,nic2}, in order  to be able to spread through the solar system, a civilisation  must be capable to build a Dyson sphere~\cite{free}, i.e.  a  maximal technological exploitation of most the energy from its local star,  which in the case of the Earth with the Sun would correspond to an energy consumption  of~$E_D\approx 4\times10^{26}$ Watts, we call this value Dyson limit. Our actual energy consumption is estimated in~$E_c\approx 10^{13}$ Watts (Statistical Review of World Energy source)~\cite{srw}. To describe our technological evolution, we may roughly schematise the development as a dichotomous random process 

\begin{equation}
\frac{d}{dt}T=\alpha T\xi(t).
\label{dic}
\end{equation}
where~$T$ is the level of technological development of human civilisation  that we can also identify with the energy consumption. ~$\alpha$  is a constant parameter describing the technological  growth rate (i.e. of $T$) and~$\xi(t)$ a random variable with values~$0,1$. We consider therefore,  {\color{black}based  on data  of global energy consumption  \cite{elov, emeritus} } an exponential growth with fluctuations mainly reflecting  changes in global economy. We therefore consider a modulated exponential growth process  where  the fluctuations  in the growth rate are captured by the variable $\xi(t)$. This variable switches between  values~$0,1$ with waiting times between switches  distributed with density~$\psi(t)$. When $\xi(t)=0$ the growth stops and resumes when $\xi$  switches to $\xi(t)=1$. If we consider $T$ more strictly as describing the technological development,   $\xi(t)$  reflects the fact that investments in research can have interruptions as a consequence of alternation of periods of economic growth and crisis. 
{\color{black}With the following transformation, 

\begin{equation}
W=\log\left (\frac{T}{T_0}\right)^{2/\alpha}-2 \langle \xi\rangle t,\,\,
\end{equation}
differentiating both sides respect to $t$ and using Eq.~(\ref{dic}), we obtain for the transformed variable $W$ 
}

\begin{equation}
\frac{d}{dt}W=\bar{\xi}(t)  
\label{dic2}
\end{equation}where 
 $\bar{\xi}(t)=2[\xi(t)-\langle \xi\rangle]$ and ~$\langle \xi\rangle$ is the average of~$\xi(t)$ so that~$\bar{\xi}(t)$ takes the values~$\pm 1$.

The above equation has been intensively studied, and a general solution for the probability distribution~$P(W,t)$ generated by a generic waiting time distribution can be found in literature~\cite{jmp}. Knowing the distribution we may evaluate the first passage time distribution in reaching the necessary level of technology to e.g.  live in  the extraterrestrial space or develop any other way to sustain population of the planet. This characteristic time has to be compared with the time that it will take to  reach the no-return point. Knowing the first passage time distribution~\cite{peter} we will be able to evaluate the probability to survive for our civilisation.

If the dichotomous process is a Poissonian process with rate $\gamma$ then the correlation function is an exponential, i.e. 

\begin{equation}\label{corr}
\langle \bar{\xi}(t)\bar{\xi}(t')\rangle = \exp[-\gamma\mid t-t'\mid]
\end{equation}
and Eq.~(\ref{dic2}) generates for the probability density the well known telegrapher's equation

\begin{equation}\label{teleg}
\frac{\partial^2}{\partial t^2}P(W,t) +\gamma\frac{\partial}{\partial t}P(W,t) =\frac{\partial^2}{\partial x^2}P(W,t) 
\end{equation}
We note that the approach that we are following is based on the assumption that at random times, exponentially distributed with rate  $\gamma$,   the  dichotomous variable $\bar{\xi}$ changes its value.
With this assumption the solution to Eq.~(\ref{teleg}) is 

\begin{equation}\label{eqc9}
P(W,t)=\frac{1}{2}\exp\left[-\frac{\gamma}{2} t \right]\left[
  \delta\left(t-\mid\!\! W\!\!\mid\right)+
 \frac{\gamma}{2}
\left(I_0\left[\frac{\gamma}{2} \sqrt{t^2-W^2}\right]+\frac{t
I_1\left[\frac{\gamma}{2} \sqrt{t^2-W^2}\right]}{
\sqrt{t^2-W^2}}\right)\theta\left(t-\mid\! W\!\mid \right)\right],
\end{equation}
where~$I_n(z)$ are the modified Bessel function of the first kind. Transforming back to the variable~$T$ we have

\begin{equation}\label{pt1}
P(T,t)=\frac{1}{4} J e^{-\frac{\gamma }{2} t}\left [\delta (2 t-x)+\delta (x)+   \gamma \left( I_0\left[\frac{\gamma}{2} \sqrt{(2 t-x) x}  \right]+\frac{  t I_1\left[\frac{\gamma}{2} \sqrt{(2 t-x) x} \right)}{\sqrt{x (2 t-x)} }\right)\right]\theta\left(t-\frac{x}{2}\right)\theta\left(x\right)
\end{equation}
where for sake of compactness we set
\begin{equation}\label{pt2}
x=\log (T/T_0)^{2/\alpha},\,\,\,\,J=\frac{dW}{dT}=\frac{2}{\alpha  T}
\end{equation}

\begin{figure}[h!]
\includegraphics[height=7.2 cm,width=5.4cm, angle=270]{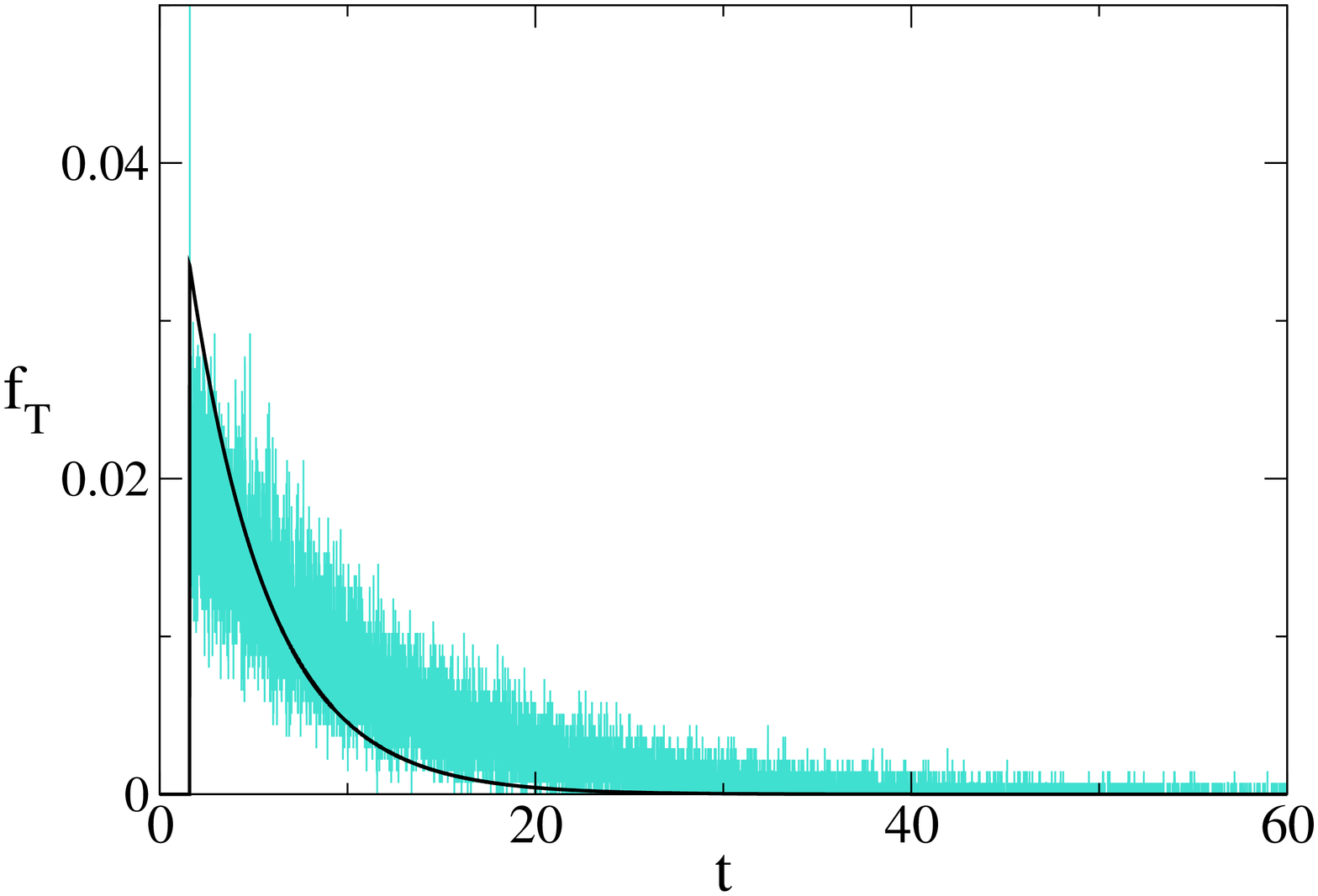}
\includegraphics[height=7.2 cm,width=5.4 cm, angle=270]{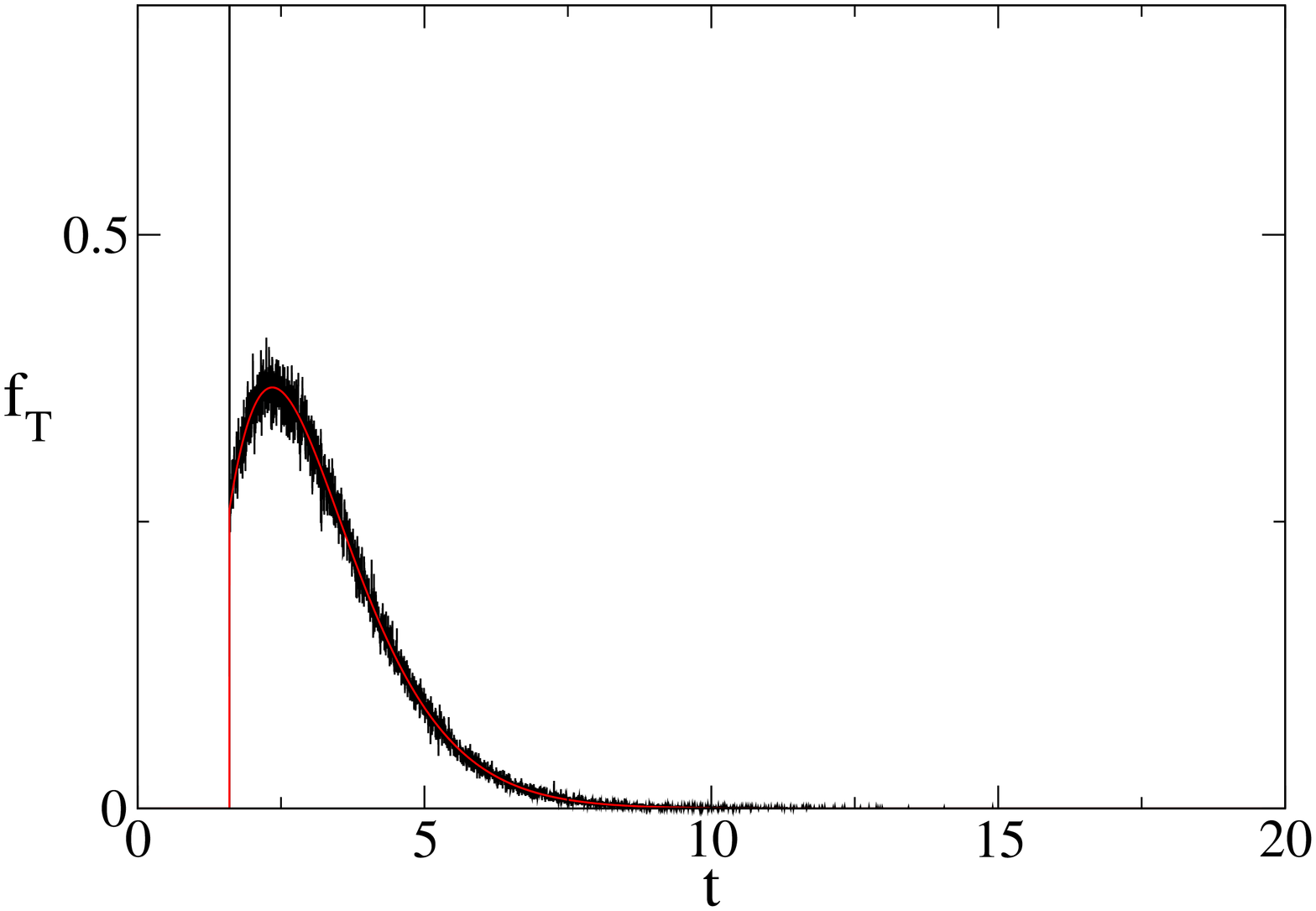}
\caption{
(Left) 
Comparison between theoretical prediction of Eq. (\ref{lapl3})  (black curve) and numerical simulation of  Eq. (3) (cyan curve) for $\gamma=4$ (arbitrary units).
(Right) Comparison between theoretical prediction of Eq. (\ref{lapl3})  (red curve) and numerical simulation of  Eq. (3) (black curve) for $\gamma=1/4$ (arbitrary units).
}
\label{fig3}
\end{figure}

\vspace{1cm}

In Laplace transform we have

\begin{equation}\label{lapl1}
\hat{P}(T,s)=\frac{J}{2 \left(\frac{\gamma }{2}+s\right)}\left[ \delta (x)+\frac{(\gamma +s)^2}{2(\frac{\gamma }{2} +  s) }
 \exp\left[-\frac{s x (\gamma +s)}{2 \left(\frac{\gamma }{2}+s\right)}\right]\right].
\end{equation}
The first passage time distribution, in laplace transform, is evaluated as~\cite{george}

\begin{equation}\label{lapl2}
\hat{f}_T(s)=\frac{ \hat{P}(x,s)}{ \hat{P}(x_1,s)}=
\exp\left[-\frac{s (\gamma +s) (x-x_1)}{2 \left(\frac{\gamma }{2}+s\right)}\right],\,\,\,\ x>x_1.
\end{equation} 
\vspace{1cm}
Inverting the Laplace transform we obtain

\begin{equation}\label{lapl3}
f_T(t)=\exp\left[-\frac{\gamma}{2} t\right] \frac{\gamma  \sqrt{x-x_1} I_1\left(\frac{\sqrt{x-x_1} \sqrt{t-\frac{x-x_1}{2}} \gamma }{\sqrt{2}}\right)}{2 \sqrt{2} \sqrt{t-\frac{x-x_1}{2}}} \theta\left[t-\frac{x-x_1}{2}\right]+ \exp\left[-\frac{\gamma}{2} t\right]\delta \left(t-\frac{x-x_1}{2}\right),
\end{equation} 
which is confirmed (see Fig. \ref{fig3})  by numerical simulations.
The time average to get the point $x$ for the first time is given by

\begin{equation}\label{tmedio}
\langle t\rangle = \int\limits_0^\infty t f_T(t)dt= 
 x-x_1=\log (T/T_0)^{2/\alpha}-\log (T_1/T_0)^{2/\alpha}=\frac{2}{\alpha} \log\left(\frac{T}{T_1} \right),
\end{equation} 
which interestingly is double the time it would take if a pure exponential growth occurred, {\color{black} depends on the ratio between final and initial value of $T$ and  is independent of $\gamma$.}
We  also stress that this result depends on parameters directly related to the stage of development of the considered civilisation, namely   the starting value $T_1$, that we  assume to be the energy consumption  $E_c $  of the fully industrialised stage of the civilisation evolution and the final value $T$, that we assume to be the Dyson  limit  $E_D $,  and the technological growth rate $\alpha$. For the latter  we may, rather optimistically, choose the value $\alpha=0.345$, following the Moore Law \cite{moore} (see next section). Using the  data above, relative to our planet's scenario, we obtain  the estimate of $\langle t\rangle\approx 180$ years.  From Figs \ref{fig1} and \ref{fig2} we see that the estimate for the no-return time are 130 and 22 years for $\beta=700$ and $\beta=170$ respectively, with the latter being the most realistic value. In either case, these estimates based on average values, being less than $180$ years, already portend not a favourable outcome  for avoiding a catastrophic collapse. Nonetheless, in order to estimate the actual probability for avoiding collapse we cannot rely on average values, but we need to evaluate the single trajectories, and count the ones that manage to reach the Dyson  limit before the 'no-return point'. We implement this numerically as explained in the following.

\subsection*{Numerical Results}
We run simulations of Eqs.~(\ref{res-inter2}), (\ref{log-inter2}) and (\ref{dic}) simultaneously for different values of of parameters~$a_0$ and~$\alpha$ for fixed~$\beta$
and we count the number of trajectories  that reach Dyson limit  before the population level reaches the "no-return point" after which rapid collapse occurs. More precisely, the evolution of $T$ is stochastic due to the  dichotomous random process $\xi(t)$, so we generate the  $T(t)$ trajectories   and at the same time we follow the evolution of the population and forest 
 density dictated by the dynamics of Eqs.~(\ref{res-inter2}), (\ref{log-inter2})~\cite{epl} until the latter dynamics reaches the no-return point (maximum in population followed by collapse). When this happens,  if the trajectory in $T(t)$ has reached the Dyson limit we count it as a success, otherwise as failure. This way we determine the probabilities and relative mean times in Figs 5,6 and 7. 
Adopting  a weak sustainability point of view our model does not specify the technological mechanism by which the successful trajectories are able to find an alternative to forests and avoid collapse, we leave this undefined and link it  exclusively  and probabilistically to the attainment of the Dyson limit.
It is important to notice that we  link the technological growth process  described by Eq.(\ref{dic})  to the economic growth and therefore we consider, for both economic and technological growth, a random sequence of growth and stagnation cycles, with mean periods of  about $1$ and $4$ years in accordance with estimates  for the driving world economy, i.e. the United States according to the National Bureau of Economic Research. \cite{NBER}. 

\begin{figure}[h!]
{
\includegraphics[height=6. cm,width=9.4 cm, angle=0]{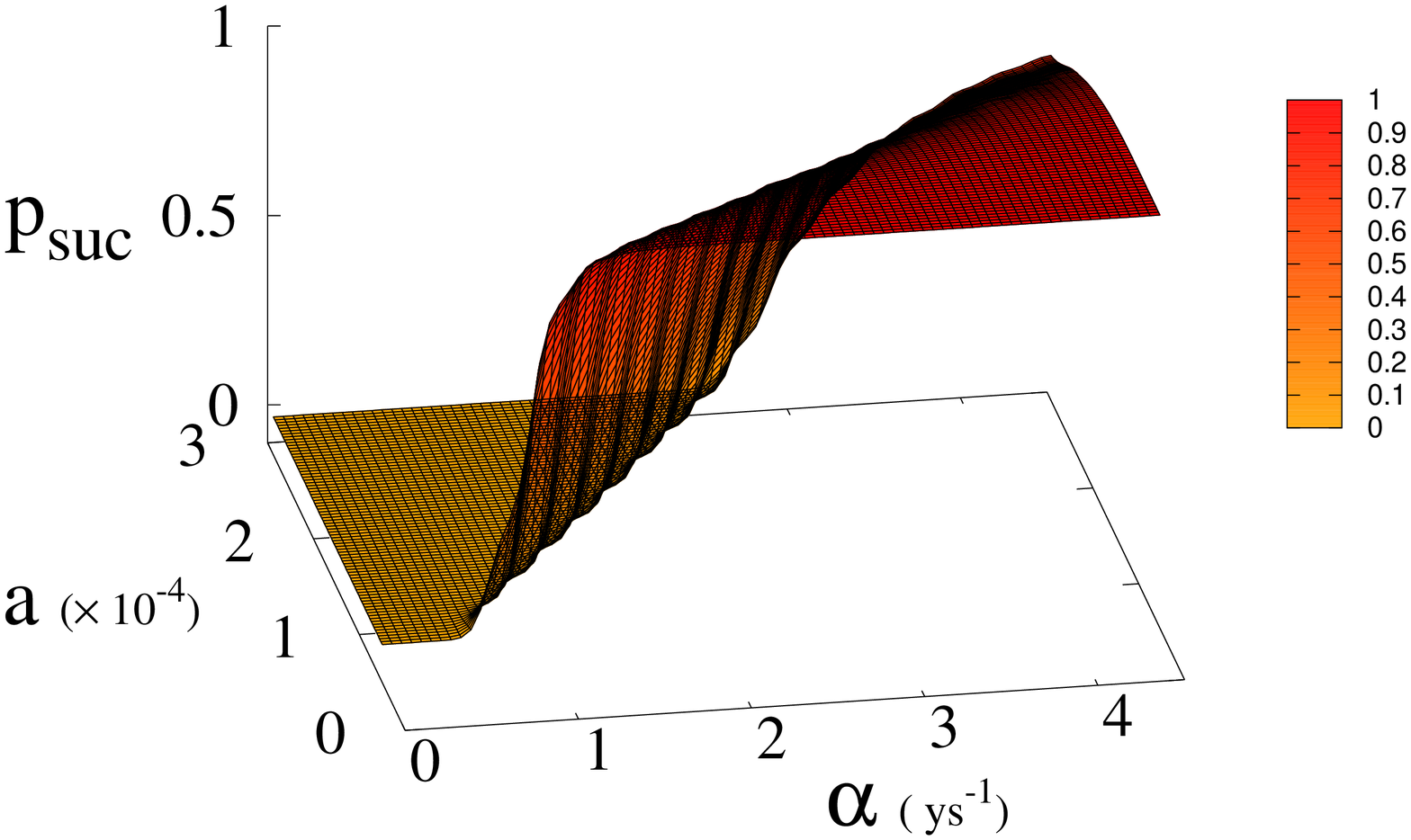}
\includegraphics[ trim=-2cm 0cm 5cm  0, height=5.5 cm,width=6.3 cm, angle=0]{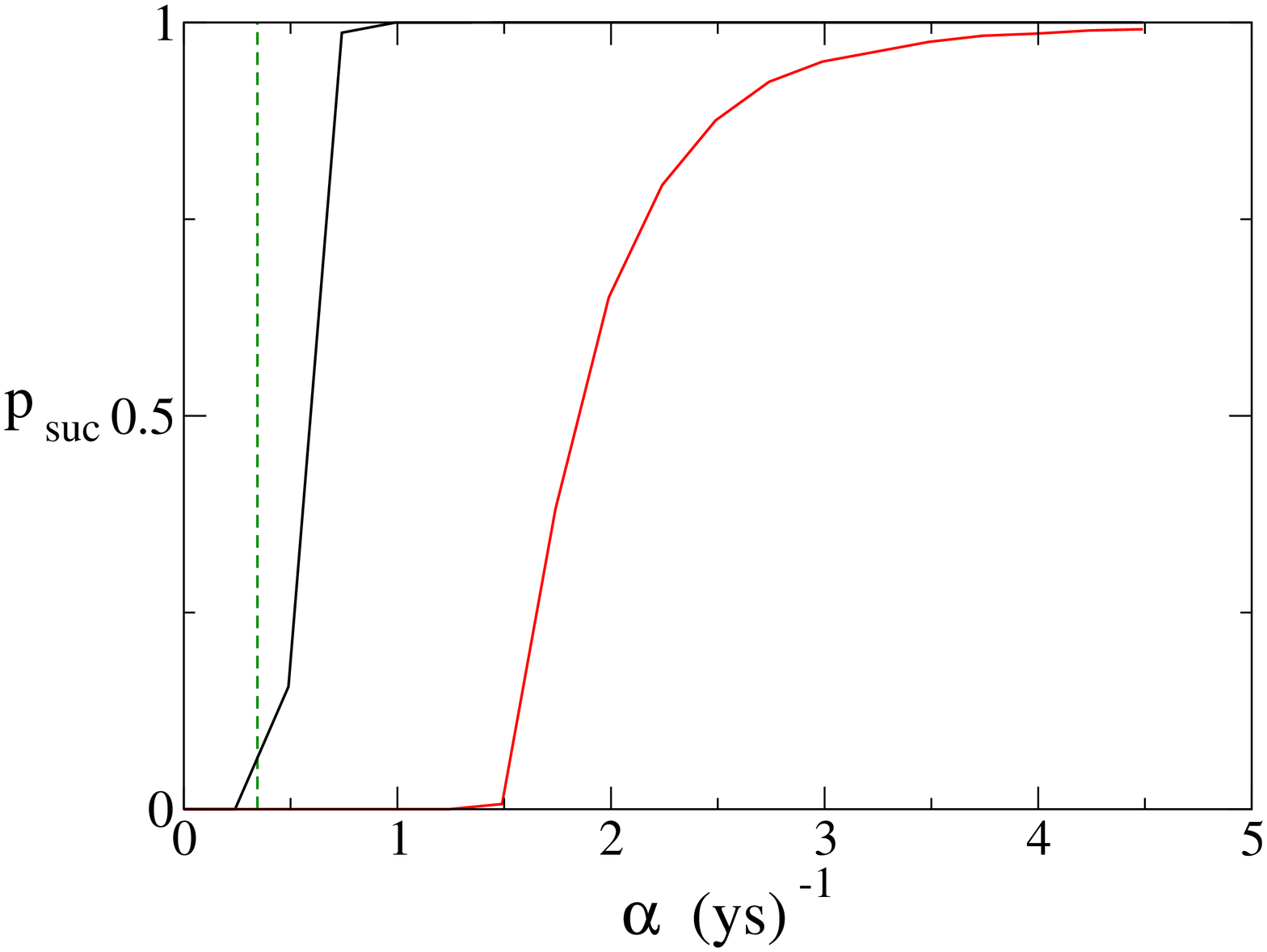}
\caption{ (Left panel) Probability $p_{suc}$ of reaching Dyson value before reaching "no-return" point as function of~$\alpha$ and~$a$ for~$\beta=170$.  Parameter $a$ is expressed in Km$^2$ys$^{-1}$.  (Right panel) 2D plot of $p_{suc}$ for  $a=1.5 \times 10^{-4}$Km$^2$ys$^{-1}$ as a function of $\alpha$. Red line is $p_{suc}$ for $\beta=170$. Black continuous lines (indistinguishable) are $p_{suc}$ for $\beta=300$ and $700$  repsectively (see also Fig \ref{fig6}).   Green dashed line indicates the value of $\alpha$ corresponding to Moore's law.  }
\label{fig4}
}
\end{figure}

In Eqs. (\ref{res-inter2}, \ref{log-inter2}) we redefine  the variables as ~$N'=N/R_W$ and~$R'=R/R_W$ with~$R_W\simeq 150 \times 10^6 Km^2$ the total continental area, and  replace parameter~$a_0$ accordingly with $a=a_0 \times R_W=1.5 \times 10^{-4} $Km$^2$ys$^{-1}$. We run simulations accordingly
starting from values  ~$R'_0$ and~$N'_0$,  based respectively  on the  current forest surface and human population. 
We take values of~$a$ from~$10^{-5}$ to~$3\times 10^{-4}$Km$^2$ys$^{-1}$ and for~$\alpha$ from~$0.01$ ys$^{-1}$ to~$4.4$ ys$^{-1}$.
Results are shown in Figs.~\ref{fig4} and ~\ref{fig6}. Figure~\ref{fig4} shows a threshold value for the parameter~$\alpha$, the technological growth rate, above which there is a non-zero probability of success. 
This threshold value increases with the value of the other parameter~$a$. As shown in Figs.~\ref{fig7} this values depends  as well on the value of $\beta$ and higher values of $\beta$ correspond to a more favourable scenario where the transition to a non-zero probability of success occurs for smaller $\alpha$, i.e. for smaller, more accessible values, of technological growth rate. More specifically,   left panel of Fig.~\ref{fig4}  shows that, for the more realistic value $\beta=170$, a region of parameter values  with non-zero probability of avoiding collapse  corresponds to values of $\alpha$ larger than 0.5. 
Even assuming  that the technological growth rate be comparable to the  value $\alpha=\log(2)/2 = 0.345$ ys$^{-1}$,  given by the Moore Law (corresponding to a doubling in size every two years), therefore,  it is unlikely  in this regime to avoid reaching the the catastrophic 'no-return point'. 
When the realistic value of $a=1.5 \times 10^{4}$ Km$^2$ys$^{-1}$ estimated from Eq.(4), is adopted, in fact,
 a probability less than $10\%$ is obtained for avoiding collapse with a Moore growth rate, even when adopting the more optimistic scenario corresponding to $\beta=700$ (black curve in right panel of Fig \ref{fig4}).  While an $\alpha$ larger than $1.5$ is needed  to have a non-zero probability of avoiding collapse when  $\beta=170$ (red curve, same panel).

\begin{figure}
\includegraphics[height=4.5 cm,width=8.6 cm, angle=0]{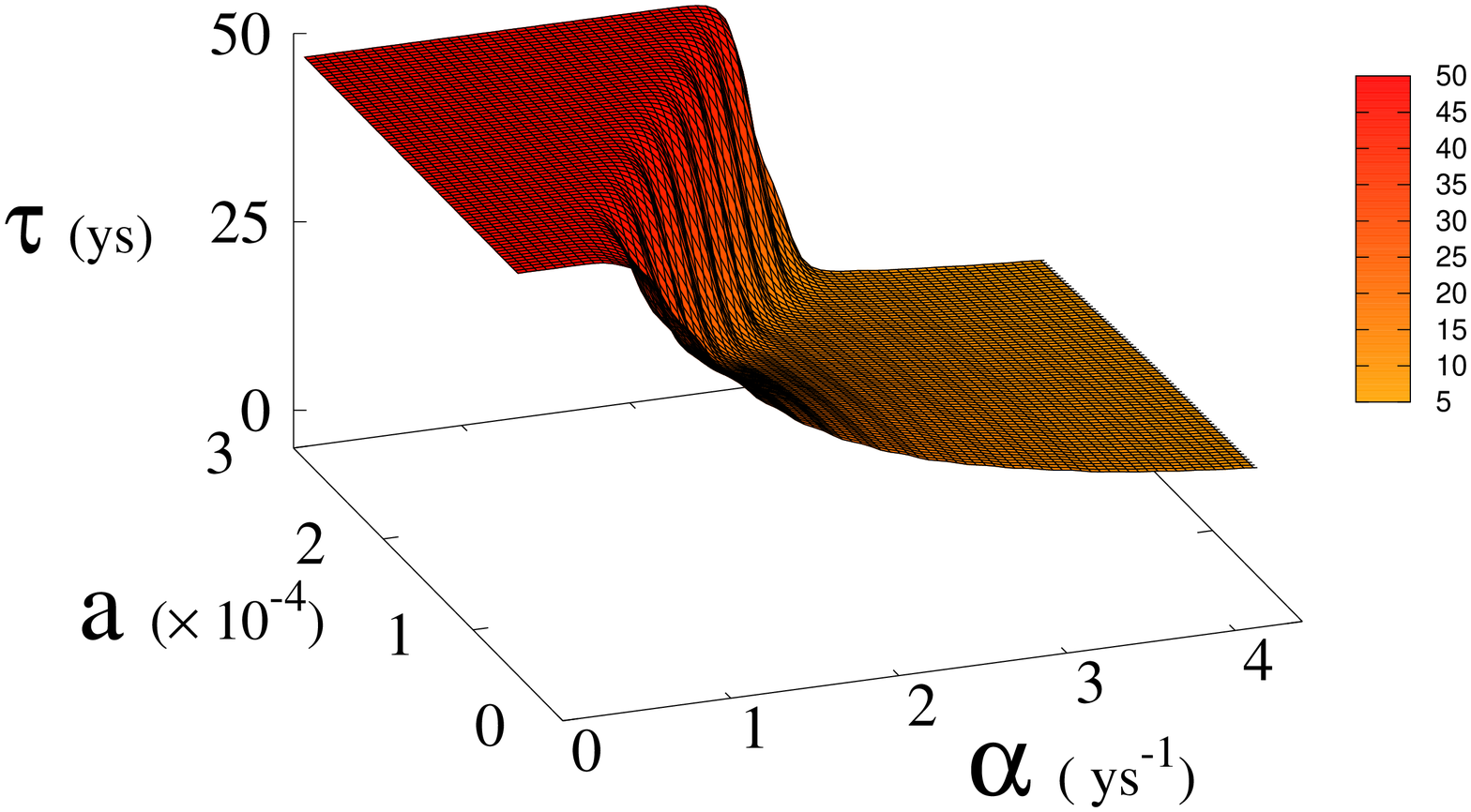}
\includegraphics[trim=-2cm 0cm 0cm  0, height=4.3 cm, width=8.6 cm, angle=0]{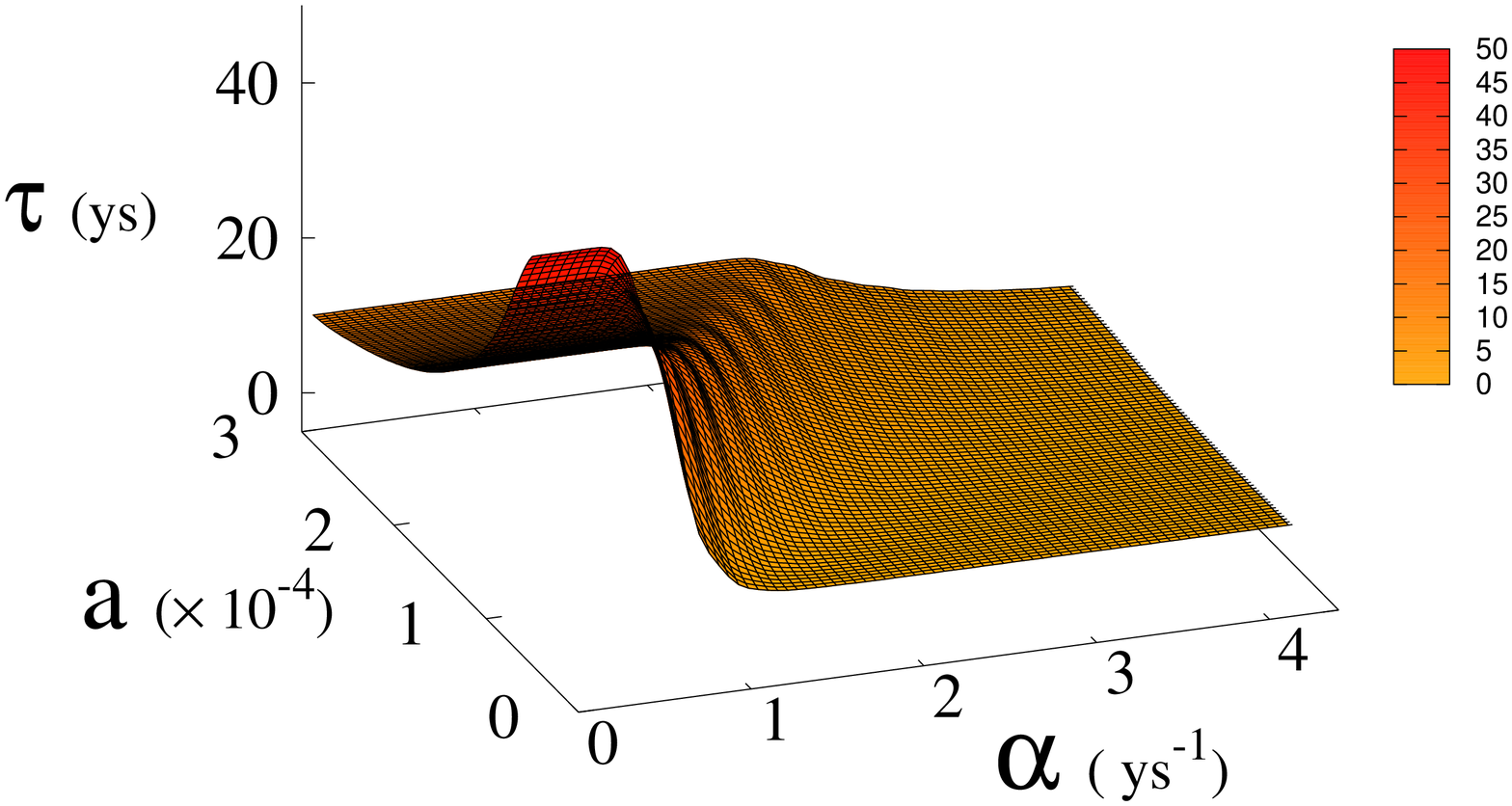}
\caption{Average time~$\tau$  (in years) to reach Dyson value before hitting "no-return" point  (success, left)  and without meeting Dyson value (failure, right) as function of~$\alpha$ and~$a$ for~$\beta=170$. Plateau region (left panel) where~$\tau \geq 50$ corresponds to diverging $\tau$,  i.e. Dyson value not being reached before hitting  "no-return" point and therefore failure.  Plauteau region at $\tau=0$ (right panel), corresponds to failure not occurring, i.e. success. Parameter $a$ is expressed in Km$^2$ys$^{-1}$. }
\label{fig5}
\end{figure}

\begin{figure}[h!]
\includegraphics[height=8. cm,width=4.8 cm, angle=270]{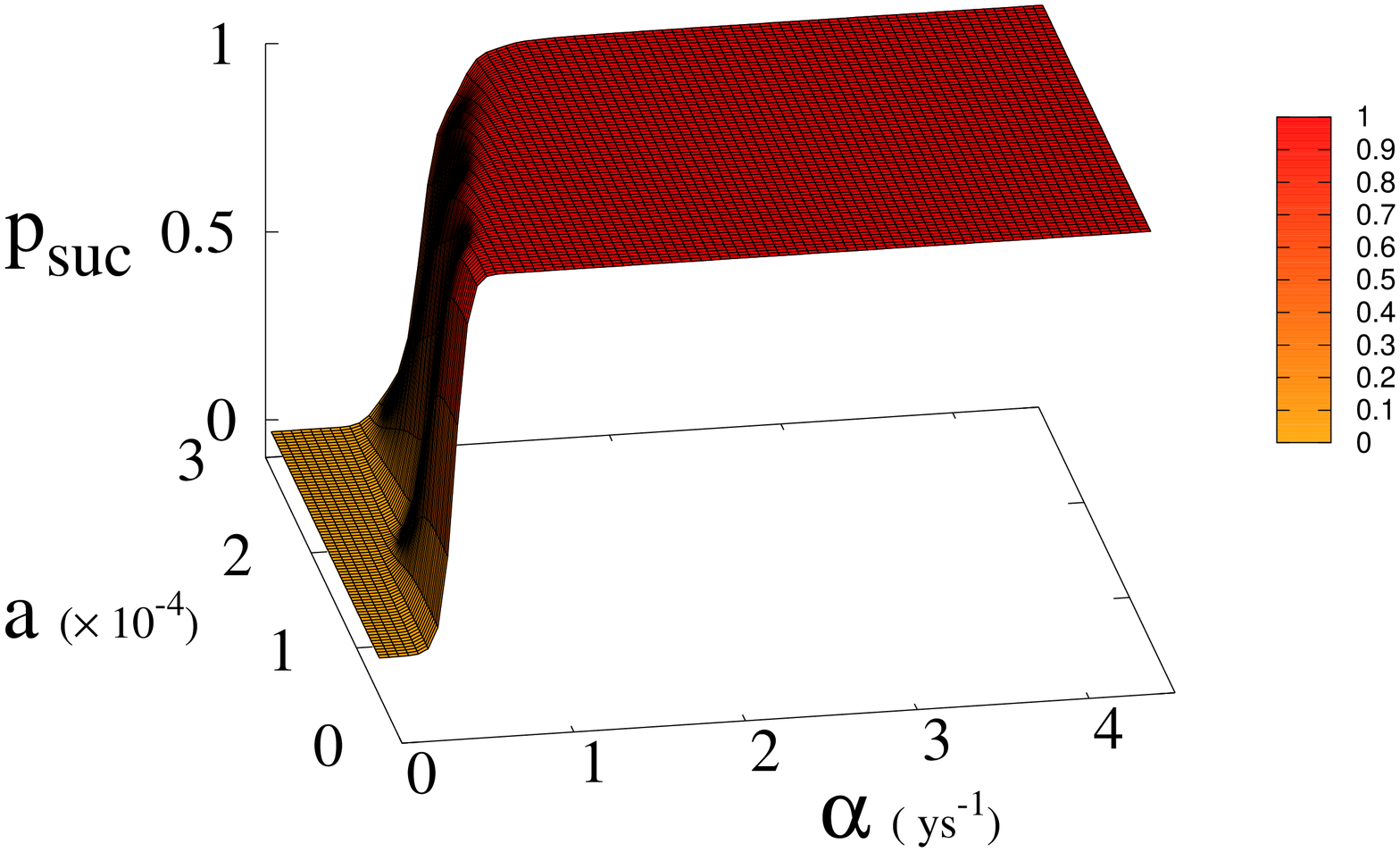}
\includegraphics[height=8. cm,width=4.8 cm, angle=270]{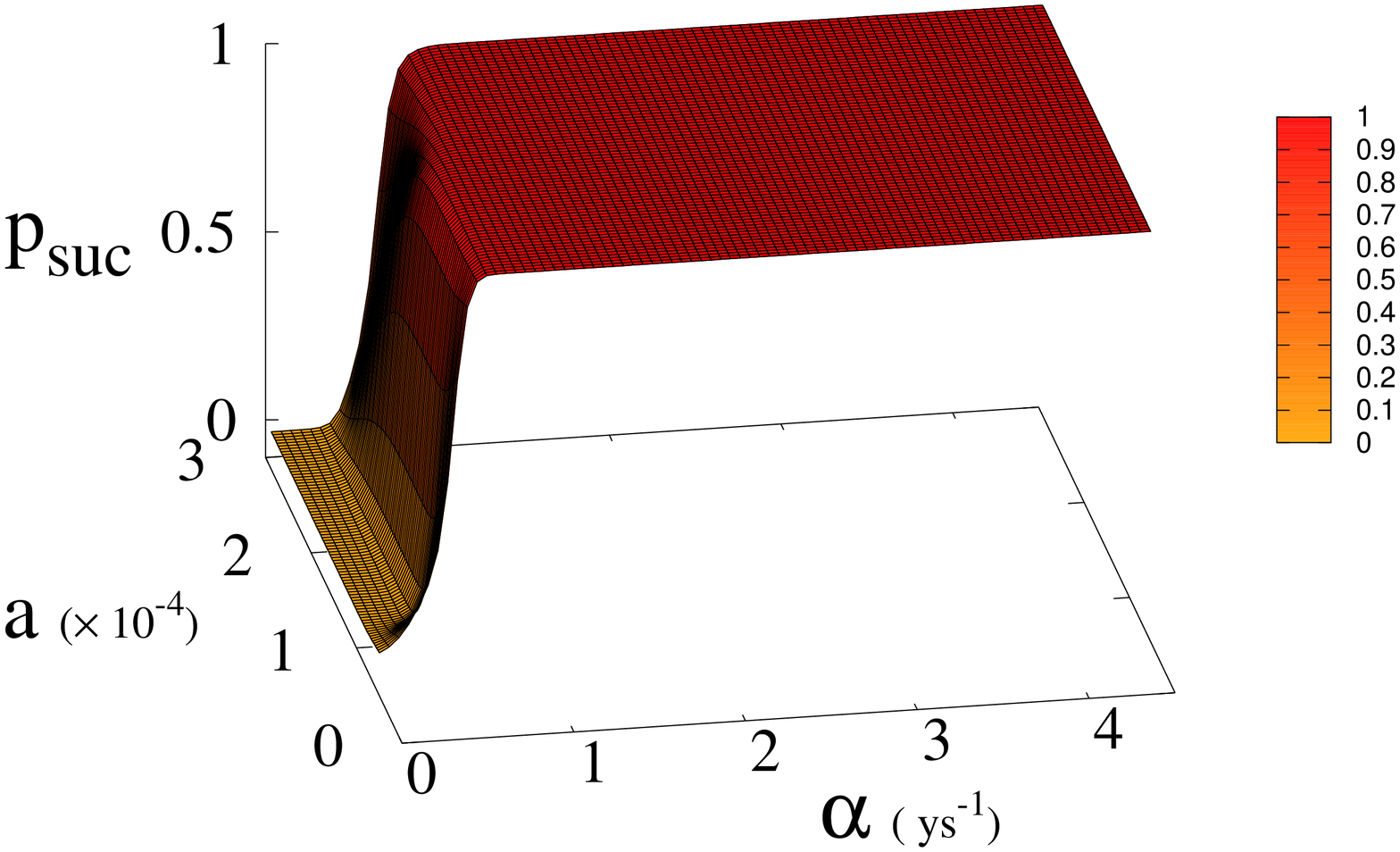}
\caption{Probability$p_{suc}$ of reaching Dyson value before hitting "no-return" point as function of~$\alpha$ and~$a$ for~$\beta=300$ (left) and $700$ (right). Parameter $a$ is expressed in Km$^2$ys$^{-1}$.}
\label{fig6}
\end{figure}

\begin{figure}[h!]
 \begin{center}
 \includegraphics[height=8. cm, width=5. cm, angle=270]{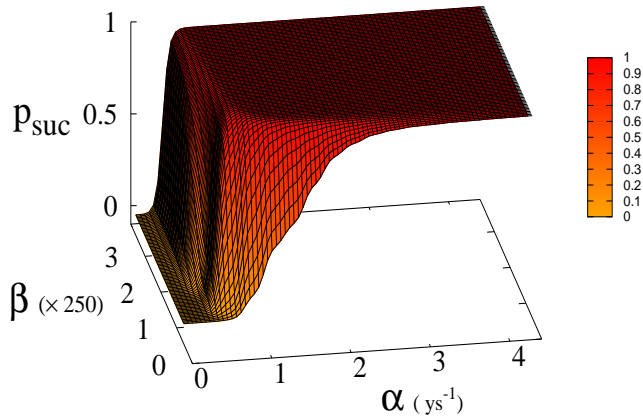}
\end{center}
\caption{Probability of reaching Dyson value~$p_{suc}$ before reaching "no-return" point as function of~$\beta$ and~$\alpha$ for~$a=1.5 \times 10^{-4} $Km$^2$ys$^{-1}$.}
\label{fig7}
\end{figure}

As far as time scales are concerned, right panel of Fig \ref{fig5} shows for $\beta=170$ that even in the range $\alpha> 0.5$, corresponding to a non-zero probability of avoiding collapse, collapse is still possible, and when this occurs, the average time to  the 'no-return point' ranges from  20 to 40 years. Left panel in same figure, shows for the same parameters, that in order to avoid catastrophe, our society has to reach the Dyson's limit in the same average amount of time of 20-40 years. 

In Fig.~\ref{fig7}  we show the dependence of the model on the parameter~$\beta$ for $a=1.5 \times 10^{-4}$. 

\section*{Methods}

We run simulations of Eqs.~(\ref{res-inter2}), (\ref{log-inter2}) and (\ref{dic})  simultaneously for different values of of parameters~$a_0$ and~$\alpha$  depending on  ~$\beta$ as explained in Methods and Results to generate Figs. 5,6 and 7.
Eqs.~(\ref{res-inter2}), (\ref{log-inter2})  are integrated via  standard Euler method. Eq. (\ref{dic}) is integrated as well via standard Euler method between the random changes of the  variable $\xi$. The 
stochastic dichotomous process $\xi$ is generated numerically in the following way: using the random number generator from gsl library we generate the  times intervals  between the changes of the dichotomous variable $\xi=0,1$, with an exponential distribution(with mean values of $1$ and $4$ years respectively), we therefore obtain a time series of $0$ and $1$ for each trajectory.   We then integrate  Eq. (\ref{dic}) in time using this time series and we average over $N=10000$ trajectories . The  latter procedure is used  to carry out simulations in Figure 3 and 4 as well in order to evaluate the first passage time probabilities. All simulations are implemented in C++.
 
\section*{Fermi Paradox}


In this section we briefly discuss a few considerations about the so called Fermi paradox  that can be drawn from our model. We may in fact relate the Fermi paradox  to the problem of resource consumption and self destruction of a civilisation. The origin of Fermi paradox dates back to a casual conversation  about extraterrestrial life that Enrico Fermi had with  E. Konopinski, E. Teller and H. York in~$1950$, during which Fermi  asked the famous question: "where is everybody?", since then become eponymous for the paradox. Starting from the closely related Drake equation~\cite{drake,burchell}, used to estimate the number of extraterrestrial civilisations in the Milky Way,  the debate around this topic has  been particularly intense in the past  (for a more comprehensive covering  we refer to Hart~\cite{hart}, Freitas~\cite{freitas} and reference therein). Hart's conclusion is that there are no other advanced or 'technological' civilisations in our galaxy as also supported recently by \cite{engler} based on  a careful reexamination of Drake's equation. In other words the terrestrial civilisation should be the only one living in the Milk Way.  Such conclusions are still debated, but many of Hart's  arguments are undoubtedly still valid while some of them need to be rediscussed or updated. For example,  there is also the  possibility that avoiding communication might actually be an 'intelligent' choice and a possible explanation of the paradox. On several public occasions, in fact, Professor Stephen Hawking suggested human kind should be very cautious about making contact with extraterrestrial life. More precisely when questioned about planet Gliese 832c's potential for alien life he once said: "One day, we might receive a signal from a planet like this, but we should be wary of answering back". Human history  has in fact been punctuated by clashes between different civilisations and cultures which should serve as caveat. From the relatively soft  replacement between Neanderthals and Homo Sapiens (Kolodny~\cite{kolo})  up to the  violent confrontation between native Americans  and Europeans, the historical examples of clashes and extinctions of cultures and civilisations  have been quite numerous. Looking at human history Hawking's suggestion appears as a wise warning and we cannot role out the possibility that extraterrestrial societies are following  similar advice coming from their best minds. 

With the help of new technologies capable of observing extrasolar planetary systems, searching and contacting alien life is becoming a concrete possibility (see for example Grimaldi~\cite{grimaldi} for a study on the chance of detecting extraterrestrial intelligence), therefore a discussion on the probability of this occurring is an important opportunity to assess also our current situation as a civilisation. Among Hart's arguments, the self-destruction hypothesis especially needs to be rediscussed at a deeper level. Self-destruction following environmental degradation is becoming more and more an alarming possibility. While violent events, such as  global war or natural catastrophic events, are of immediate concern to everyone, a relatively slow consumption of the planetary resources may be not perceived as strongly as a mortal danger for the human civilisation.  Modern societies are in fact driven by Economy, and, without giving here a well detailed definition of  "economical society",  we may agree that  such a kind of society privileges the interest of its components with less or no concern  for the whole ecosystem that hosts them (for more details see \cite{ecologicaleconomics} for a review on Ecological Economics and its criticisms to mainstream Economics) . Clear examples of the consequences of this type of societies  are the international agreements about  Climate Change. The Paris climate agreement~\cite{uni,tol} is in fact, just the last example of a weak agreement  due to its strong subordination to the economic interests of the single individual countries. In contraposition to this type of society we may have to redefine a different model of society,  a "cultural society", that in some way privileges the interest of the ecosystem above the individual interest of its components, but eventually in accordance with the overall communal interest.   
This consideration suggests a statistical explanation of Fermi paradox: even if  intelligent life forms were very common (in agreement  with the mediocrity principle in one of its version~\cite{rood}: "there is nothing special about the solar system and the planet Earth") only very few civilisations would be able to reach a sufficient technological level so as to spread in their own solar system before collapsing  due to resource consumption. 

We are aware that several objections can be raised against this argument and we discuss  below the one that we believe to be the most important.
The main objection is that we do not know anything about extraterrestrial life. Consequently, we do not know the role that a hypothetical intelligence plays in the ecosystem of the planet. For example not necessarily the planet needs trees (or the equivalent of trees) for its ecosystem. Furthermore the intelligent form of life could be itself the analogous of our trees, so avoiding the problem of the "deforestation" (or its analogous). But if we assume that we are not an exception (mediocrity principle) then independently of the structure of the alien ecosystem, the intelligent life form would exploit every kind of resources, from rocks to organic resources  (animal/vegetal/etc), evolving towards a critical situation. Even if we are at the beginning  of the extrasolar planetology, we have strong indications that Earth-like planets have the volume magnitude of the order of our planet. In other words, the resources that alien civilisations have at their disposal are, as order of magnitude,  the same for all of them, including ourselves. Furthermore the mean time to reach the Dyson limit as derived in Eq. 6  depends only on the ratio between final and initial value of $T$ and therefore would be independent of the size of the planet, if we assume as a proxy for $T$ energy consumption (which scales with the size of the planet), producing a rather general  result which can be extended to other civilisations.  Along this line of thinking, if we are an exception in the Universe we have a high probability  to collapse or become extinct, 
while if we assume the mediocrity principle  we are led to conclude that very few civilisations are able to reach a sufficient technological level so as to spread in their own solar system before the consumption of their planet's resources  triggers a catastrophic population collapse. 
The mediocrity principle has been questioned (see for example Kukla~\cite{andr} for a critical discussion about it)  but on the other hand the idea that the humankind is in some way "special" in the universe has historically been challenged several times. Starting with the idea of the Earth at the centre of the universe (geocentrism), then  of the solar system as centre of the universe (Heliocentrism) and finally our galaxy as centre of the universe. All these beliefs have been denied by the facts. Our discussion, being focused on the resource consumption, shows that whether we assume the mediocrity principle or our "uniqueness" as an intelligent species in the universe, the conclusion does not change. Giving a very broad  meaning  to the concept of cultural civilisation as a civilisation not strongly ruled by economy, we suggest  for avoiding collapse \cite{strunz} that only civilisations capable of such a  switch from an economical society to a sort of "cultural" society  in a timely manner, may survive.  This discussion leads us to the conclusion that, even assuming the mediocrity principle, the answer to "Where is everybody?" could be a lugubrious " (almost) everyone is dead".

\section*{Conclusions}
{\color{black}In conclusion our model shows that a catastrophic collapse in human population, due to resource consumption, is  the most likely scenario of the dynamical evolution  based on current parameters.  Adopting a combined deterministic and stochastic model we conclude from a statistical point of view that the probability that our civilisation survives itself is less than $10\%$ in the most optimistic scenario.  } Calculations show that, maintaining the actual rate of population growth and resource consumption, in particular forest consumption, we have a few decades left before an irreversible collapse of our civilisation (see Fig \ref{fig5}). Making the situation even worse, we stress once again  that it is unrealistic to think that the decline of the population in a situation of strong environmental degradation would be a non-chaotic and well-ordered decline. This consideration leads to an even shorter remaining time.  {\color{black} Admittedly, in our analysis, we assume  parameters  such as population growth and deforestation rate in our model as constant. This is a rough approximation  which allows us to predict future scenarios based on current conditions. Nonetheless the resulting mean-times for a catastrophic outcome to occur, which are of the order of  2-4 decades (see Fig. \ref{fig5}) ,  make this approximation  acceptable, as it is hard to imagine, in absence of very strong collective efforts, big changes of these parameters to occur in such time scale.} This interval of time seems to be out of our reach and incompatible with the actual rate of the resource consumption on Earth, although some fluctuations  around this trend are possible  \cite{natureoptimistic}   not only due to unforeseen effects of climate change but  also to desirable human-driven reforestation.  
This scenario offers as well a plausible additional explanation  to the fact that no signals from other civilisations are detected. {\color{black} In  fact  according to Eq. (\ref{tmedio})  the mean time to reach Dyson sphere depends on the ratio of the technological level $T$ and  therefore, assuming energy consumption (which scales with the size of the planet) as a proxy for $T$, such ratio is  approximately independent of the size of the planet. Based on this observation and on the mediocrity principle,   one could  extend the results shown in this paper,  and conclude that a  generic civilisation has approximatively two centuries starting from its fully developed industrial age to reach the capability to spread through its own solar system.} In fact,  giving a very broad  meaning  to the concept of cultural civilisation as a civilisation not strongly ruled by economy, we suggest that only civilisations capable of a  switch from an economical society to a sort of "cultural" society  in a timely manner, may survive.  




\section*{Acknowledgements (not compulsory)}

M.B. and G.A. acknowledge Phy.C.A for logistical support.

\section*{Author contributions statement}

M.B. and G.A. equally contributed and reviewed the manuscript.

\section*{Additional information}

\textbf{Accession codes} (where applicable); \textbf{Competing interests} The authors declare no competing interests. 


\end{document}